\documentclass[sigconf,nonacm]{acmart}
\usepackage{fancybox}
\AtBeginDocument{%
  \providecommand\BibTeX{{%
    \normalfont B\kern-0.5em{\scshape i\kern-0.25em b}\kern-0.8em\TeX}}}

\copyrightyear{2021}
\acmYear{2021}
\setcopyright{acmcopyright}
\acmConference[CIKM '21]{Proceedings of the 30th ACM International Conference on Information and Knowledge Management}{November 1--5, 2021}{Virtual Event, QLD, Australia}
\acmBooktitle{Proceedings of the 30th ACM International Conference on Information and Knowledge Management (CIKM '21), November 1--5, 2021, Virtual Event, QLD, Australia}
\acmPrice{15.00}
\acmISBN{978-1-4503-8446-9/21/11}
\acmDOI{10.1145/3459637.3481989}

\acmSubmissionID{de3124}

\begin{document}
\fancyhead{}

\title{Form 10-Q Itemization}

\author{Yanci Zhang}
\authornote{Additional information and sample dataset could be requested.}
\email{yanci@wharton.upenn.edu}
\affiliation{%
  \institution{Wharton Research Data Services (WRDS), The Wharton School}
  \streetaddress{3819 Chestnut Street – Suite 300}
  \city{Philadelphia}
  \state{PA}
  \country{USA}
  \postcode{19104}
}

\author{Tianming Du}
\affiliation{%
 \institution{Penn Medicine}
 \institution{University of Pennsylvania}
 \city{Philadelphia}
 \state{PA}
 \country{USA}
}

\author{Yujie Sun}
\affiliation{%
  \institution{Wharton Research Data Services}
  \city{Philadelphia}
  \state{PA}
  \country{USA}
}

\author{Lawrence Donohue}
\affiliation{%
 \institution{Wharton Research Data Services}
 \city{Philadelphia}
 \state{PA}
 \country{USA}
}

\author{Rui Dai}
\authornotemark[1]
\email{rdai@wharton.upenn.edu}
\affiliation{%
  \institution{Wharton Research Data Services}
  \streetaddress{3819 Chestnut Street – Suite 300}
  \city{Philadelphia}
  \state{PA}
  \country{USA}
  \postcode{19104}
}


\begin{abstract}
The quarterly financial statement, or Form 10-Q, is one of the most frequently required filings for US public companies to disclose financial and other important business information. 
Due to the massive volume of 10-Q filings and the enormous variations in the reporting format, it has been a long-standing challenge to retrieve item-specific information from 10-Q filings that lack machine-readable hierarchy. 
This paper presents a solution for itemizing 10-Q files by complementing a rule-based algorithm with a Convolutional Neural Network (CNN) image classifier. This solution demonstrates a pipeline that can be generalized to a rapid data retrieval solution among a large volume of textual data using only typographic items.
The extracted textual data can be used as unlabeled content-specific data to train transformer models (e.g., BERT) or fit into various field-focus natural language processing (NLP) applications.
\end{abstract}



\maketitle

\section{Introduction}

Publicly traded firms in the United States are required to file financial statements and other disclosure documents that provide a comprehensive review of the firm’s business operations and financial condition to the Securities and Exchange Commission (SEC). In turn, the SEC makes many of those filings publicly available to improve financial transparency and reduce fraud. 

Disclosure files could be in plain text or as an HTML file designed for human ingestion. However, these documents lack the hierarchical structure seen in JSON or XML files, optimized for machine extraction \cite{zhifeng_xml}\cite{zhifeng_search}. Additionally, although following a predefined reporting structure, companies have considerable flexibility in changing the tagging, wording, and presentation of these documents. As a result, the same types of SEC filings may vary significantly between companies and over time, prohibiting researchers, regulators, and even many well-equipped practitioners from utilizing machine power to extract information \cite{cohen_lazy_2020}\cite{foucault_corporate_2019}\cite{li_annual_2008} . The purpose of this study is to present a systematic solution for retrieving data from the quarterly financial statement, refereed as the Form 10-Q, which is one of the most critical financial disclosures to the SEC. In particular, we commit to developing a system capable of efficiently retrieving typographically formatted data from 10-Qs that lack a machine-readable hierarchy. The output would be item-level records in key-value storage database.

A fundamental challenge to information retrieval from 10-Qs is the considerable amount and variety of its reporting formats. By 2019, about 587,000 10-Qs have been submitted to the SEC by over 26,000 companies. The structure of most 10-Qs, primarily in HTML format, is only required to visually comply with the SEC's typographic structure specification, arranged in the $\boldsymbol{Part}$s and $\boldsymbol{Item}$s, as illustrated in Table~\ref{tab:layout}. As a result, information extraction would rely heavily on keyword searching, such as "Item 1. Legal Proceedings", to parse the whole text into its constituent items. However, the substantial volume of highly tailored 10-Qs significantly complicates this effort. The following is a list of the most often encountered troublesome scenarios.

\begin{figure*}[ht]
  \centering
  \shadowsize=1mm
    \color{gray}
    \shadowbox{\fboxsep=0.1mm\fcolorbox{white}{white}{
    \includegraphics[width=0.9\textwidth]{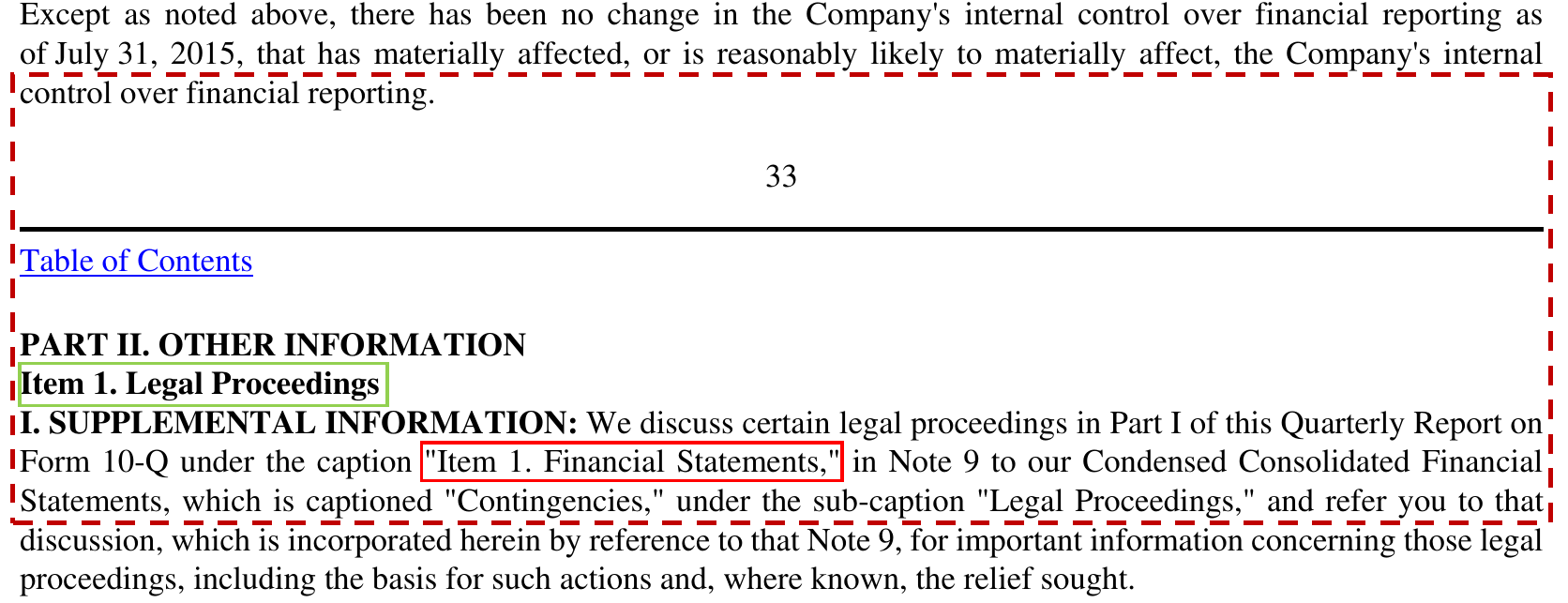}}
    }
  \caption{
  A portion of documentation sample. Green rectangle shows the title of an $\boldsymbol{Item}$, which is the beginning of the target content. Red rectangle shows the $\boldsymbol{Item}$ mentioned as reference in a paragraph. Red dot rectangle shows an input to CNN.
  }
  \label{fig:example}
\end{figure*}

\begin{enumerate}
\item Each company has its tailored design of HTML presentations, which could vary quarterly. The layout of a 10-Q may not align with the typographical structure defined by SEC.
\item Keywords are not universal - companies may not use the exact wording from SEC's predefined structure, like "Item 1. Legal Proceedings" in Table~\ref{tab:layout} but instead, use similar titles like "Item i" or "Legal Proceedings."
\item Keywords can be present throughout the main text without necessarily being the title of the Item or Part. Those keywords frequently appear in other places like a table of content, paragraph references, page break, or even referenced in a paragraph, as shown in Figure~\ref{fig:example}.
\end{enumerate}
The difficulties outlined above will inevitably result in either time-consuming case-by-case manual collection or an omission of a significant portion of 10-Q files, resulting in bias and inconsistent analytic results. Many previous studies take the latter approach, presumably due to the magnitude of collection burdens. For instance, a literature \cite{loughran_when_2011} reported a similar task on 10-K, annual disclosure to the SEC, at a rate of around 75\%, while subsequent work \cite{dyer_evolution_2017} record an even lower rate at a similar effort.  

\begin{figure*}[ht]
  \centering
  \includegraphics[width=\textwidth]{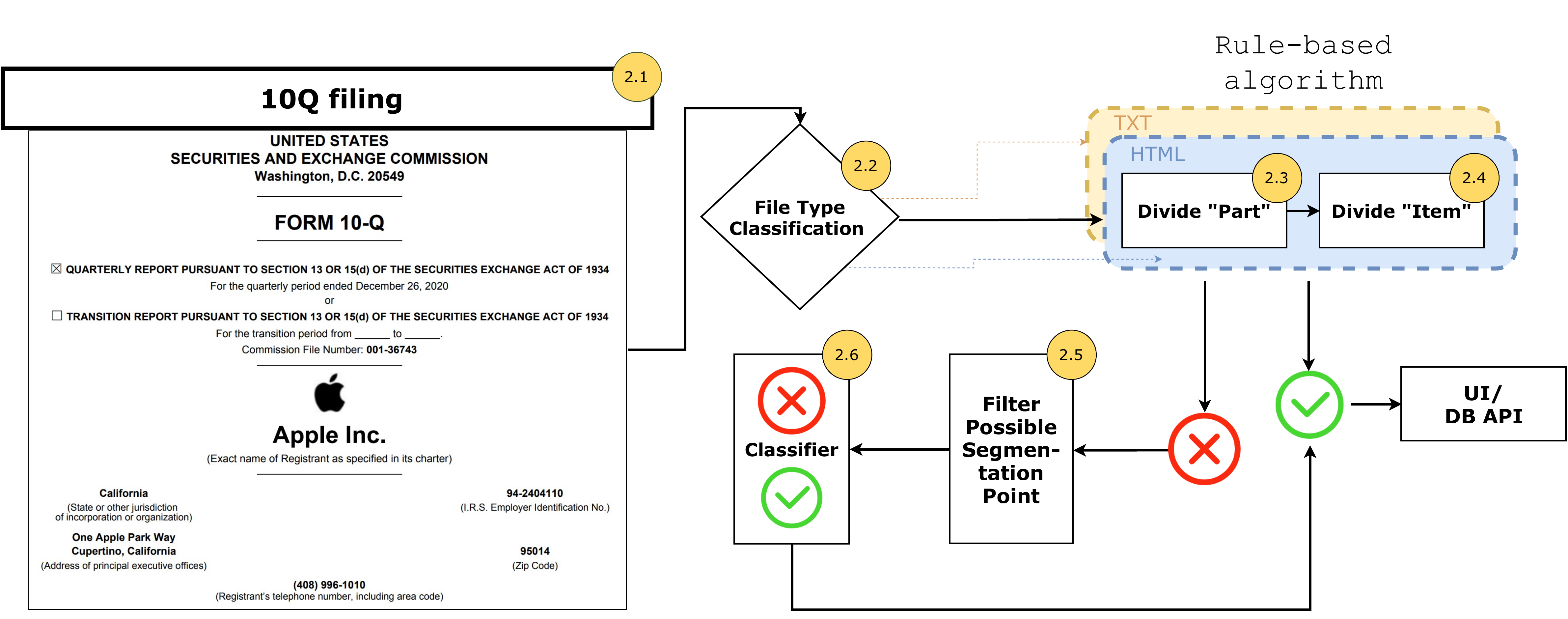}
  \caption{The pipeline design diagram. Yellow circles include references to the corresponding section. Dashed rectangular box represents slightly different algorithms for two data types. Green check mark and red cross indicates the process success or not.
  The cover page of Apple Inc. 10-Q \cite{aapl} is used as an example.
  }
  \label{fig:pipeline}
\end{figure*}

Given that the past literature has not yet provided a generalized and reliable methodology for resolving such challenges, our study presents a viable solution using multi-stage processes via a pipeline method. First, it employs a rule-based approach to extract textual information from the majority of 10-Qs. Then, the remainder of 10-Q information is extracted using a Convolutional Neural Network (CNN) image classifier trained on visual characters applied by human reviewers. A graphical presentation illustrates the flow of our method (Figure~\ref{fig:pipeline}), and each component on the pipeline will be addressed in detail in the following subsections.

Our pipeline methodology can accommodate a wide range of 10-Q variances, most notably in the HTML layouts used by various companies and reporting periods. Performance-wise, the pipeline converts 10-Q documents to a machine-readable format within 0.1 seconds. In summary, the pipeline processes 10-Q documents available on the SEC website in HTML or plain text format\footnote{Database available at  \url{https://www.sec.gov/edgar/searchedgar/companysearch.html}} and generates hierarchical textual outputs.  Each item in a 10-Q is dissected into separate text files, designated in their respective section as \{document id\}\_\{Part number\}\_\{Item number\}. Our demo tool
\footnote{Additional information, demo tool, and sample dataset could be requested.} 
interface is illustrated in Figure~\ref{fig:website}.


This pipeline's output would enable a variety of downstream tasks, including non-quantitative natural language processing challenges related to asset pricing in company risk and management discussion 
\cite{bao_simultaneously_2014,bank_qilin,NBERw28444,network,graph,DAI2020,zhu2020high,jarrow2021low,zhu2021time,zhu2020adaptive,zhu2021news,zhu2021clustering,Jie2021BiddingVC,pmlr-v48-la16,PrashanthL2016CumulativePT,jiecheng-thesis,ZHAO2018619,LIN20181,jiecheng8329994,huang2020time,huang2021staying,begel2020lessons,yingjie_capittal,leshang_pricing,leshang9101712,Guo2019UnderreactionOA,Fang2020InsuranceAL,mao2014pasa,harvey2017edos,zheng2019flowcon,zheng2019target,acharya2019workload,wang2014fresh,mao2017draps,fu2019progress,chen2020woa,cheng2021network,yang2021optimizing,chen2021improved,bo2021subspace,bo2019novel,xie2018data,haoyugraph,haoyusemantic,eventtrigger,CARPENTER2021679,yunzi_forcast,zhiqiang_8k}, which are becoming increasingly attractive issues in finance and accounting research.


\begin{table} 
\caption{The SEC 10-Q Layout}
\label{tab:layout}
\begin{tabular}{l}
\hline \hline
\textbf{Part I --- Financial Information} \\ \hline
Item 1. Financial Statements. \\
Item 2. MD\&A Condition and Results of Operations. \\
Item 3. Quantitative and Qualitative Disclosures \\
\hspace{.38in}About Market Risk. \\
Item 4. Controls and Procedures. \\ \hline
\textbf{Part II --- Other Information \hspace{1.5in}} \\ \hline
Item 1. Legal Proceedings. \\
Item 1-A. Risk Factors. \\
Item 2. Unregistered Sales of Equity Securities and Use of \\
\hspace{.37in} Proceeds. \\
Item 3. Defaults Upon Senior Securities. \\
Item 4. Mine Safety Disclosures. \\
Item 5. Other Information. \\
Item 6. Exhibits. \\  \hline  \hline
\end{tabular}
\end{table}

\section{Pipeline Design}
The pipeline depicted in Figure~\ref{fig:pipeline} is composed of its constituent elements. A rule-based algorithm and a CNN image classifier are two critical components. As discussed in Sections 2.3 and 2.4, the rule-based approach successfully retrieves 95\% of our 10-Q sample. Additionally, when paired with other pipeline elements, the CNN image classifier can handle the 4\% of our 10-Q sample that our rule-based algorithm cannot, as stated in Section 2.6.

\subsection{Structure of 10-Q document}
Form 10-Q is a quarterly financial disclosure mandated by the SEC. The SEC gives precise instructions on filing formats, including 2 $\boldsymbol{Part}$s and 11 $\boldsymbol{Item}$s covering financial and other information to enhance companies' transparency to shareholders and regulators. In Table~\ref{tab:layout}, we supply those predefined titles for the $\boldsymbol{Part}$s and $\boldsymbol{Item}$s.

\subsection{File Type Classification}
The SEC 10-Q filing typically has two types of formats – HTML and text. Most filings in text format are antiquated, and more recent filings from 2003 onwards are predominantly in HTML format. Though data of both formats are available to be processed, this paper mainly focuses on the HTML format given that recent and up-to-date filings are relevant ones to be examined, and HTML also covers more diverse situations. A similar algorithm equivalent to section 2.3 to process unstructured plain text files is also implemented. A classifier is designed to identify the data type, and send the data into algorithms with their corresponding input format.

\subsection{Identify Parts and Parse}
Due to the structure constraint described previously, we use a divide-and-conquer approach to organize and decompose the textual contents using standard 10-Q layouts. This strategy consists of two steps: the first is to divide a document into two $\boldsymbol{Part}$s; the second is to divide each $\boldsymbol{Part}$ into $\boldsymbol{Item}$s and handle a smaller group of paragraphs. This two-step method is critical for 10-Q, as its item numbering system contains repeated $\boldsymbol{Item}$ numbers by design (for example, each $\boldsymbol{Part}$ shares some identical title, "Item 1"). Without segmenting 10-Q by $\boldsymbol{Part}$s, the keyword-based dividing procedure may result in some unexpected errors.

Due to the fact that the majority of those 10-Qs adhere to SEC guideline, though loosely, we separate them into 10-Q $\boldsymbol{Part}$s using a rule-based algorithm. We include several mechanisms for detecting and resolving mistakes at various levels of the method for $\boldsymbol{Part}$ division. This section of the algorithm requires three pillars to interpret the 10-Q filings at the $\boldsymbol{Part}$s level,

\begin{enumerate}
\item The first pillar utilizes the hyperlink provided in the table of content to locate the separation points between two  $\boldsymbol{Part}$s. 
\item The next pillar uses regular expressions to accomplish the same operation when errors occur in the hyperlink approach.
\item The final pillar utilized the first 10-Q Part II or equivalent titles that appeared in the HTML header of new pages if both earlier pillars are failed to obtain meaningful records. 
\end{enumerate}
These three pillars ensured that, when errors are detected in an earlier stage, a specific algorithm corresponding to the error would be invoked as an attempt to bypass the error and retrieve the contents.

\subsection{Identify Items Sequentially}
Assuming content are successfully divided at $\boldsymbol{Part}$s level, there is no duplicate $\boldsymbol{Item}$s under the same $\boldsymbol{Part}$, so it can be reliably accomplished using a single pass sequential process. 

However, there are several typical scenarios in which the $\boldsymbol{Part}$ or $\boldsymbol{Item}$ titles can be falsely detected. Figure~\ref{fig:example} illustrates a good example. The green rectangle could be considered as a subtitle under Part II, rather than an $\boldsymbol{Item}$ title. Other negative cases include but are not limited to appearing in the table of content, titles, paragraph references (red rectangle), or page breaks.

Despite the existence of those challenges, till this step, this rule-based algorithm can still achieve high accuracy for more than 95.43\% of 10-Q filings of our selected 10-Q samples, as shown in the result section. We will then focus on coping with the remaining cases in the following two subsections. 

Finally, it is worth highlighting the importance of this procedure. The U.S. stock market can simultaneously comprise more than 8,000 public companies at a given time, 5\% of which may result in up to 400 unreadable 10-Qs in a quarter.

\subsection{Segmentation Point Identification}
To resolve the 5\% inaccuracy generated by the rule-based algorithm, we resort to filtering out all potential segmentation points by exact keywords matching and generating corresponding context data for further classification. Unfortunately, the keywords searching step are not used directly at the early stage of our pipeline because Item (Part)-related keywords may exist multiple locations in a 10-Q, resulting in a significant increase in the computing time for CNN image classifier. Thus, we trade off time-intensity with a smaller set of challenging cases to improve our overall performance.

\subsection{Identify Items through Classification}
To identify Items, we used image-based CNN methods \cite{yancy9175642,du2021adaptive,du2021training,du2019morphology,du2015algorithm,du2018real,liu2019detection,shi2020uenet,zhai2019coronary,reza2017label,sanet,sefd,dvrnet,headmotion,Le_2020_WACV,shilei9091528,shilei8293810,Josh8695110,Xu2020ImprovedSB,Xu2019UNetWO,Gao2020ADNETAN,Hu2019AccuratePM,Xu2020SmallBD,tang2021spatial,tang2019clinically,tang2019nodulenet,tang2018automated,tang2021recurrent,li2021frequentnet}
to address this classification problem. The typographic structure in 10-Q may be more readily captured through aesthetic qualities, especially when many text-based models, such as NLP methods, commonly ignore font size, case, and text position on a page. As a result, we change this textual analysis problem to a computer vision issue to determine the document's structure. In Figure~\ref{fig:example}, a sample input to CNN is depicted in the red dot line rectangle box. Our approach adopts a binary classification using a ResNet34 \cite{resnet} architecture.

In addition, other baseline approaches are tested. 
In terms of model parameter, 8 closest neighbors and inverse-distance as the weight factor are for k-nearest neighbors model. For Decision Tree model, the depth limits to maximum 8 and weak classifiers cap in AdaBoost to be 100. 
Features are chosen from our massive observations, applied as following: 1. A binary value equal to one if an $\boldsymbol{Item}$ \{Item number\}.\{Item name\} is in bold, 0 otherwise. 2. A binary value equal to one if an $\boldsymbol{Item}$ \{Item number\}.\{Item name\} is aligned in the center, 0 otherwise 3. The number of space characters on the left of $\boldsymbol{Item}$ \{Item number\}.\{Item name\}. 4. The number of space characters on the right of $\boldsymbol{Item}$ \{Item number\}.\{Item name\}. 5. The number of characters included in $\boldsymbol{Item}$ \{Item number\}.\{Item name\}.

\section{Experiment and Results}
Our data usage, performance and results will be presented in this section. We will first spell out the performance of our first half of the pipeline, defined as processes before 2.5, followed by the second half performance, defined as processes after 2.5, and eventually summarizing the whole data operation process indicated in Figure~\ref{fig:pipeline}. 

\subsection{First Half of Pipeline Performance}

Our rule-based algorithm processed altogether 193,000 documents, belonging to 8,100 unique firms published between 1994 and 2017, reporting time-out errors which fails to locate two $\boldsymbol{Part}$s for 1,246 (0.65\%) filings and fails to parse Part 1 for 4,038, and Part 2 for 3,557 filings respectively. These errors indicate a 4.57\% algorithm limitation for which a postprocess will be applied on. 

To verify the validity and accuracy of our rule-based model, we randomly selected 5,570 $\boldsymbol{Item}$s including 3,528 $\boldsymbol{Item}$s that are considered correct itemizations and another 2,042 ones that are considered false, defined in 2.3 and shown Figure~\ref{fig:example}. We asked three human labelers for manual check - in particular, they were required to determine whether an $\boldsymbol{Item}$ is correctly captured as the title (accurate parsing) instead of as words inside the content paragraph or anywhere else (inaccurate parsing). Within all the labeled data, there are 2,880 companies covered, each company has an average of 1.5 filings covered in this set over the same period. No duplicate of company is removed given format changes over quarters. 

As shown in Table~\ref{tab:labelers}, the precision for the 95.43\% portion claimed to be correctly retrieved is 99.72\%, which yields an itemization performance of 95.16\%. This alone beats the performance of previous literature \cite{loughran_when_2011} at 75\% and our replication of it at 70\%.

\subsection{Second Half of Pipeline Performance}
Although our rule-based approach outperformed the previous baseline, it is nowhere near a perfect algorithm. The machine learning models and CNN are tested and proven to be crucial as a building block (section 2.6) in our integrated pipeline. Our final implementation used a CNN model, which is a slightly modified version of ResNet34 at the last layer. 

In terms of labelled data, there are two separate sets. First coming from verification process described in section 3.1, were used to train our models. The train, validation and test split are 8:1:1. Train and test results are presented. Challenging sample is defined as files initially failed in 4.57\% mentioned in section 3.1. Second, for this challenging portion of data, because they are not included in previous set, our coauthor team manually labeled the data ourselves. Within 1000 total data points, we have 371 positive and 629 negative observations. 

As can be seen in Table~\ref{tab:train_res}, the CNN model yields the best result, without additional rules or human generated features. We consider it as a successful attempt of converting textual analysis problem into a computer vision problem to identify the document’s hierarchy. 

Nevertheless, it is worth noting that the trained CNN model is not designed to be deployed independently without our rule-base method because of the enormous possibility of $\boldsymbol{Item}$s identified by keyword search described in section 2.5. Our stand-alone CNN model would significantly deteriorate pipeline performance in speed and accuracy (e.g., 83.4\% in challenging cases). As the first half of the pipeline ensures both accuracy and speed, the CNN model performs the best as a building block to solve challenging cases in the whole system. 

The overall scientific performance of the whole system would be around 98.97\% (95.16\% + 4.57\% × 0.834), or approximately a thousand filings in total require human attention. Our actual large-scale result on over half million filings has a 1.6\% failure rate.

\begin{table} 
\caption{Human label verification for Section 2.4 Rule-based}
\label{tab:labelers}
\begin{tabular}{cccl}
\hline \hline
& Labeled True & Labeled False & Total \\ \hline
Rule-Based True&3,518&10&3,528\\  
Rule-Based False&7&2,035&2,042\\  
Total&3,525&2,045&5,570\\
\hline  \hline
\end{tabular}
\end{table}

\begin{table}
\caption{Model performance for pipeline Section 2.6. Data of Panel A \& B are from Table~\ref{tab:labelers}, Panel C is from 4.57\% labeled.}
\label{tab:train_res}
\begin{tabular}{lcccc}
\hline \hline 
&	Recall	&	Precision	&	Accuracy	&	$F_1$	\\\hline
\multicolumn{5}{c}{\emph{Panel A:Training Sample}}\\ \hline 
CNN	&	0.997	&	0.997	&	0.998	&	0.997	\\
Logistic Regression	&	0.911	&	0.923	&	0.897	&	0.917	\\
Naive Bayes	&	0.922	&	0.909	&	0.894	&	0.916	\\
SVM	&	0.922	&	0.912	&	0.896	&	0.917	\\
$k$-Nearest Neighbor&	0.960	&	0.955	&	0.946	&	0.957	\\
Decision Tree	&	0.960	&	0.956	&	0.947	&	0.958	\\
AdaBoost	&	0.952	&	0.938	&	0.931	&	0.945	\\\hline 
\multicolumn{5}{c}{\emph{Panel B: Testing Sample}}\\ \hline 
CNN	&	0.967	&	0.983	&	0.984	&	0.975	\\
Logistic Regression	&	0.977	&	0.947	&	0.935	&	0.962	\\
Naive Bayes	&	0.930	&	0.968	&	0.916	&	0.949	\\
SVM	&	0.938	&	0.953	&	0.909	&	0.945	\\
$k$-Nearest Neighbor	&	0.946	&	0.992	&	0.948	&	0.968	\\
Decision Tree	&	0.930	&	0.992	&	0.935	&	0.960	\\
AdaBoost	&	0.977	&	0.940	&	0.929	&	0.958\\ \hline 
\multicolumn{5}{c}{\emph{Panel C: Challenging Sample}}\\ \hline 
CNN	&	0.841	&	0.834	&	0.879	&	0.837	\\
Logistic Regression	&	0.776	&	0.780	&	0.836	&	0.778	\\
Naive Bayes	&	0.803	&	0.772	&	0.839	&	0.787	\\
SVM	&	0.798	&	0.769	&	0.836	&	0.783	\\
$k$-Nearest Neighbor	&	0.671	&	0.798	&	0.815	&	0.729	\\
Decision Tree	&	0.650	&	0.785	&	0.804	&	0.711	\\
AdaBoost	&	0.806	&	0.782	&	0.845	&	0.794	\\
\hline  \hline
\end{tabular}
\end{table}

\section{Demonstration Interface}
\begin{figure}[ht]
  \centering
  \shadowsize=1mm
    \color{gray}
    \shadowbox{\fboxsep=0.1mm\fcolorbox{white}{white}{
    \includegraphics[width=0.95\linewidth]{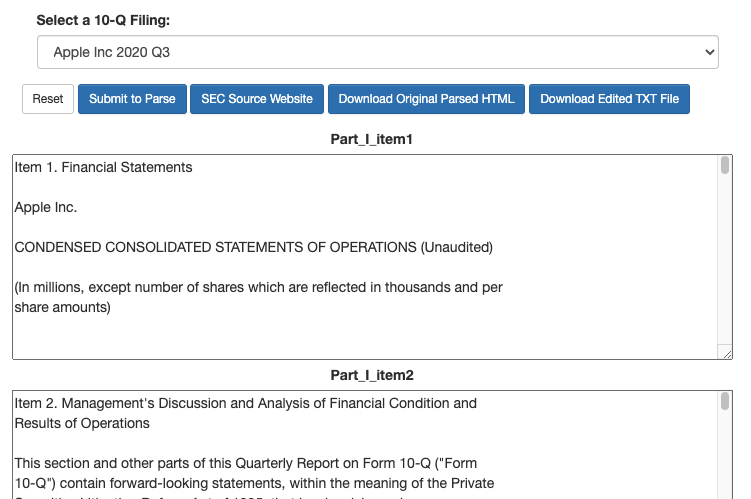}}
    }
  \caption{
  Interface to allow itemization, edits and downloads
  }
  \label{fig:website}
\end{figure}

Figure \ref{fig:website} presents a user interface for researchers and analysts to quickly identify desired $\boldsymbol{Item}$s information; users can have a view of the itemized content, edit the data in text boxes and export them in different formats. In addition to the demonstration to accelerate single document research, we would show another demo example of downstream application using the information generated by our proposed system at a large scale.

\section{Conclusion}
In this paper, we purposed an integrated pipeline to retrieve itemized information from 10-Qs. It provides a solution for the long-standing problem of large-scale itemization for finance practitioners and researchers. We believe it would enable further research in risk analysis, management strategy and asset pricing by leveraging massive amount of data retrieved from this work.

\clearpage
\bibliographystyle{ACM-Reference-Format}
\bibliography{citations}


\end{document}